\documentclass[sigconf, authorversion]{acmart}

\usepackage{subfig}

\AtBeginDocument{%
  }

\setcopyright{acmlicensed}
\copyrightyear{2023}
\acmYear{2023}
\acmDOI{10.1145/3603719.3603733}

\acmConference[SSDBM 2023]{35th International Conference on Scientific and Statistical Database Management}{July 10--12, 2023}{Los Angeles, CA, USA}

\acmPrice{15.00}
\acmISBN{979-8-4007-0746-9/23/07}

\begin{document}

\title[Selecting Efficient Cluster Resources for Data Analytics]{Selecting Efficient Cluster Resources for Data Analytics:\\When and How to Allocate for In-Memory Processing?}

\author{Jonathan Will}
\email{will@tu-berlin.de}
\affiliation{%
    \institution{Technische Universität Berlin}
    \city{Berlin}
    \country{Germany}
}

\author{Lauritz Thamsen}
\email{lauritz.thamsen@glasgow.ac.uk}
\affiliation{%
    \institution{University of Glasgow}
    \city{Glasgow}
    \country{United Kingdom}
}

\author{Dominik Scheinert}
\email{dominik.scheinert@tu-berlin.de}
\affiliation{%
    \institution{Technische Universität Berlin}
    \city{Berlin}
    \country{Germany}
}

\author{Odej Kao}
\email{odej.kao@tu-berlin.de}
\affiliation{%
    \institution{Technische Universität Berlin}
    \city{Berlin}
    \country{Germany}
}

\renewcommand{\shortauthors}{Will et al.}

\begin{abstract}
Distributed dataflow systems such as Apache Spark or Apache Flink enable parallel, in-memory data processing on large clusters of commodity hardware.
Consequently, the appropriate amount of memory to allocate to the cluster is a crucial consideration.

In this paper, we analyze the challenge of efficient resource allocation for distributed data processing, focusing on memory.
We~emphasize that in-memory processing with in-memory data processing frameworks can undermine resource efficiency.
Based on the findings of our trace data analysis, we compile requirements towards an automated solution for efficient cluster resource \mbox{allocation}.

\vfill  %

\end{abstract}




\maketitle

\vfill\null %
\section{Introduction}\label{sec:INTRODUCTION}

Distributed dataflow systems such as Apache Spark~\cite{zaharia2010spark} and Apache Flink~\cite{carbone2015apache} enable parallel data processing on large clusters of commodity hardware by facilitating error handling and parallelization.

\vspace{4.5mm} %

\newpage

However, choosing the appropriate cluster resources to allocate to such data processing jobs is difficult~\cite{lama2012aroma, rajan2016perforator}, especially in public clouds, where the user is virtually unrestricted in the amount of resources available.  %
Some choices, such as the total number of CPU cores allocated, generally represent a trade-off between lower cost or faster execution, which is a matter of user preference.
Meanwhile, misallocations of other resources, especially the amount of memory, can lead to significant bottlenecks or underutilization.
Such inefficiencies can multiply the cost of dataflow jobs executed on public cloud resources~\cite{alipourfard2017cherrypick, hsu2018arrow, al2022juggler, al2022blink, will2022get}.

\begin{figure}[h!]
    \centering
    \subfloat{%
      \includegraphics[width=.5\linewidth]{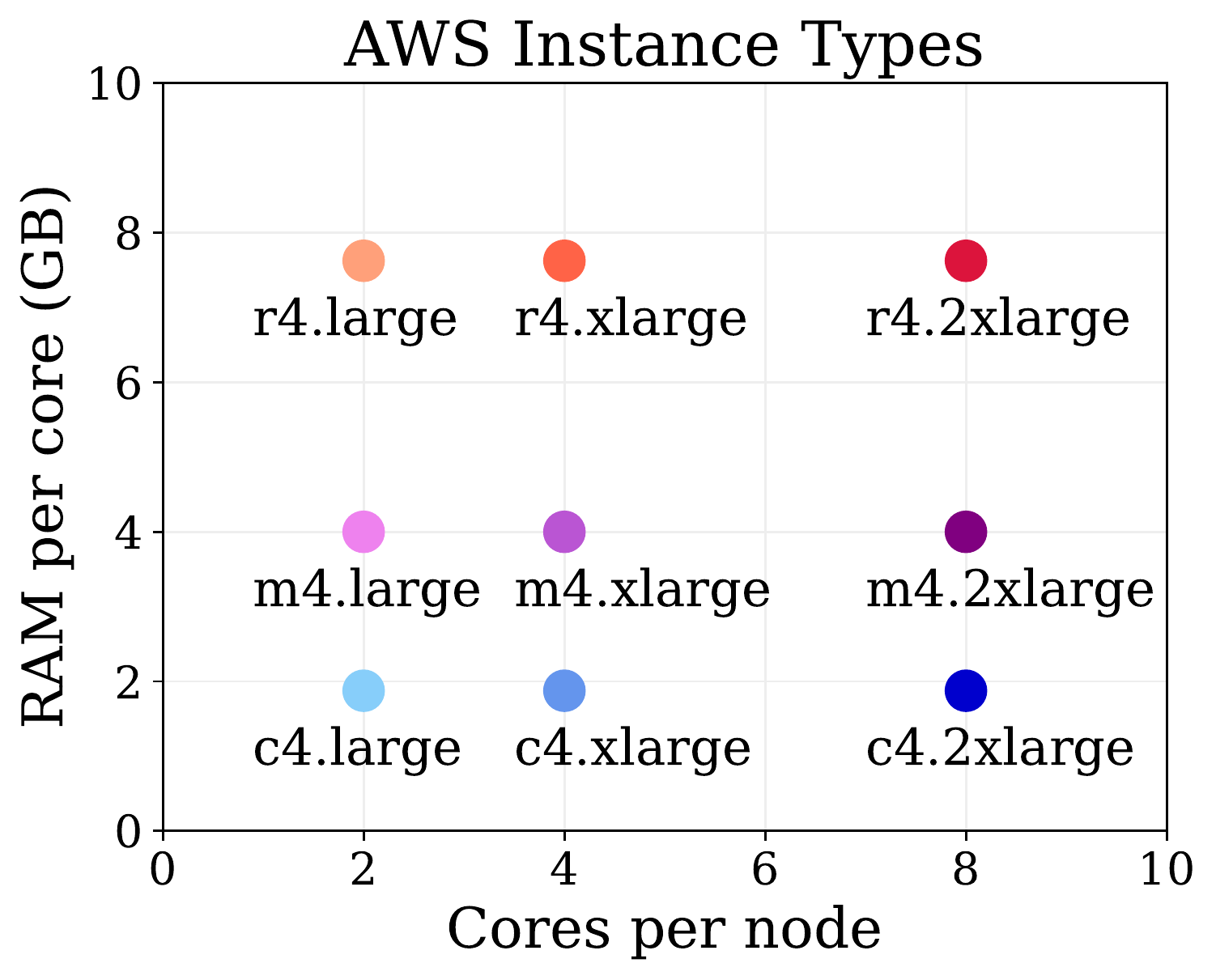}%
    }\hfill
    \subfloat{%
      \includegraphics[width=.5\linewidth]{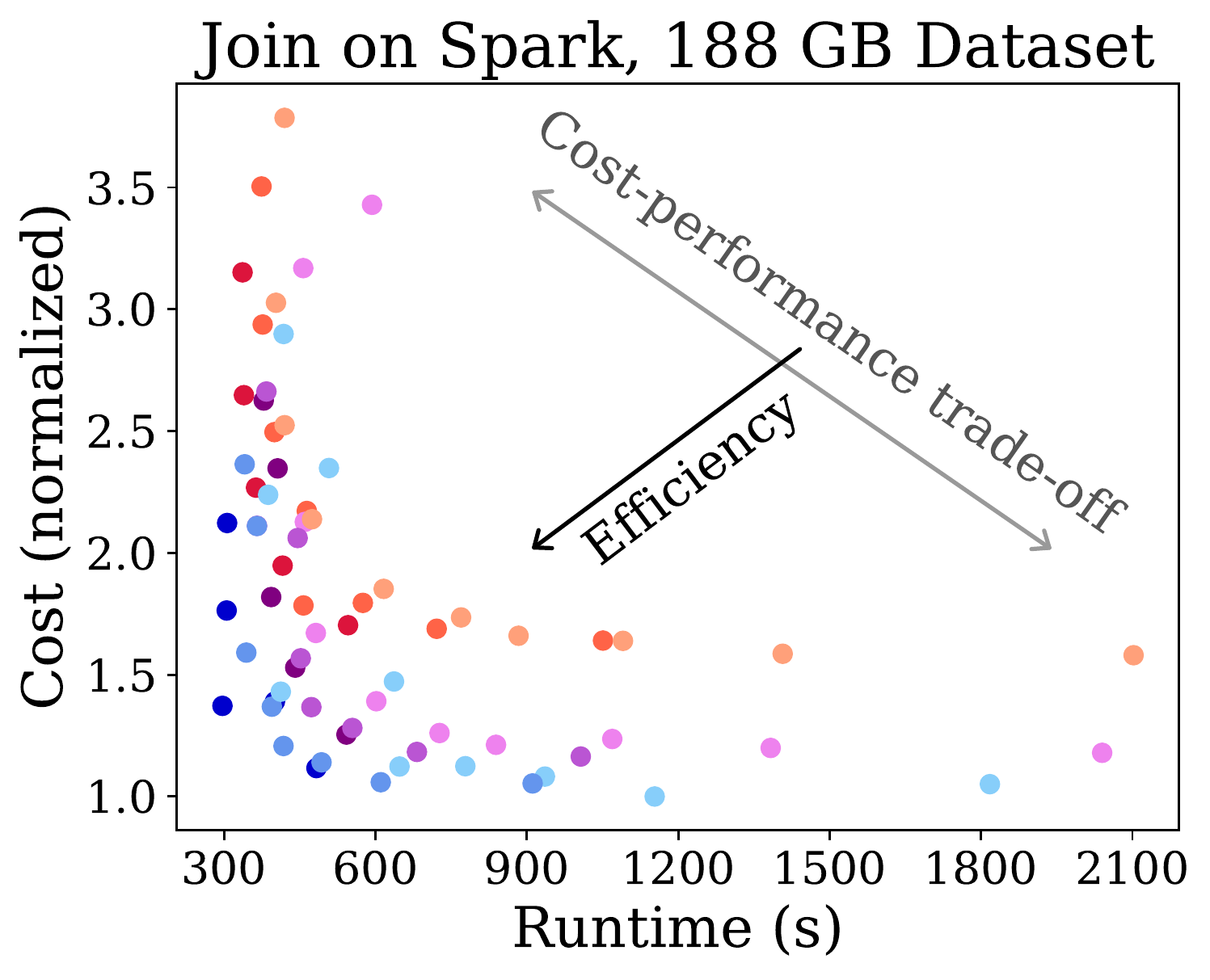}%
    }\hfill
    \vspace{-4mm}
    \caption{Resource efficiency and cost-performance trade-off.}\label{fig:one}
    \Description{Resource efficiency and cost-performance trade-off.}
\end{figure}

The concepts of the cost-performance trade-off and resource efficiency are visualized in Figure~\ref{fig:one}.
The example we see is a Spark join job, executed on clusters ranging from 4 to 48 nodes, consisting of VMs of one out of nine different Amazon Web Services (AWS) instance types.
The data is from a publicly available trace dataset, which is described in more detail in Section~\ref{sec:METHODOLOGY}.

\vspace{-4.5mm}
\vspace{-2mm}

Many methods for assisting users in resource selection incur significant overhead to collect training data for performance models~\cite{venkataraman2016ernest, thamsen2016bell, alipourfard2017cherrypick, hsu2018arrow, chen2021silhouette, scheinert2021potential}.
In comparison, several recent approaches employ lightweight profiling runs on reduced data with reduced hardware to measure the memory footprint of the dataset relative to its size.~\cite{will2022get, will2022ruya, al2022juggler, al2022blink}.
This allows them to predict memory usage in a full run and then allocate just enough memory to the cluster to facilitate in-memory processing.
However, these approaches fail when the cost of allocating enough memory for in-memory processing outweighs the benefit of avoiding disk I/O operations (from spilling the cache from memory to disk).
They also fail when allocating enough memory to cache the whole dataset forces public cloud users to then also allocate more CPU cores that do not increase performance meaningfully.
This depends on the scale-out behavior of the given job.

\clearpage

In this paper, we argue that, despite significant progress, important challenges remain in the allocation of memory and other resources.
We illustrate these challenges with a problem analysis.
Then, we discuss the difficulties of efficiently gathering enough information to make an informed resource allocation decision, and finally, we summarize the requirements for a solution.

\section{Methodology}\label{sec:METHODOLOGY}

This section formulates our research questions and limits the scope of the problem by defining the constraints.
It then provides an overview of the trace dataset used for the problem analysis in Section~\ref{sec:ANALYSIS}.

\subsection{Research Questions}

This paper aims to provide preliminary results towards answering the following research questions:

\newcommand{\C}[1]{\texttt{C}\hspace{.21mm}\texttt{#1}}
\newcommand{\RQ}[1]{\texttt{RQ}\hspace{.21mm}\texttt{#1}}

\begin{itemize}
    \item \RQ{1}: When is it resource-efficient to allocate enough memory to a cluster to facilitate in-memory processing?\\\vspace{-3mm}
    \item \RQ{2}: How to design automated resource allocation approaches that can allocate the right amount of memory and other resources without relying on costly trained models?
\end{itemize}

\subsection{Scope and Limitations}

In defining the scope, we set the following constraints:

\begin{itemize}
    \item \C{1} \emph{- Distributed batch processing}:\\
        We limit the scope to distributed batch processing applications of distributed dataflow systems.\\\vspace{-3mm}
    \item \C{2} \emph{- Initial configuration}:\\
        We only look for a static resource configuration and do not attempt to change it at runtime.\\\vspace{-3mm}
    \item \C{3} \emph{- Focus on memory, CPU cores, and scale-out}:\\
        We limit the configuration options to a variable number of VMs with varying amounts of memory and CPU cores.\\\vspace{-3mm}
    \item \C{4} \emph{- Unlimited resources}:\\
        We assume the availability of virtually unlimited resources, as is realistic in public clouds or large on-premises clusters.
\end{itemize}

\subsection{Trace Dataset}  %

For the problem analysis, we used a popular public trace dataset\footnote{\href{https://github.com/oxhead/scout}{github.com/oxhead/scout}, accessed in April 2023.} which was introduced by Hsu et al.\ in \emph{Arrow}~\cite{hsu2018arrow}.
The dataset contains 1031 unique Spark and Hadoop MapReduce~\cite{dean2008mapreduce} executions, facilitated by the benchmarking tool HiBench by Intel.
There are eight different underlying algorithms and each job has been executed with two different dataset sizes.

The jobs were run on 69 different AWS cluster configurations.
The cluster configurations have scale-outs between 4 and 48 VMs and these have instance types of classes \emph{c}, \emph{m}, and \emph{r} in the sizes \emph{large}, \emph{xlarge}, and \emph{2xlarge}.
Virtual machines of type \emph{c} have less memory per core than those of type \emph{r}, while those of type \emph{m} fall between the two.
The denominations \emph{large}, \emph{xlarge}, and \emph{2xlarge} refer to the number of cores per machine.
The instance types and their exact resource configurations are shown in Figure~\ref{fig:one}, the color coding of which is used in all subsequent figures.

\section{Problem Analysis}\label{sec:ANALYSIS}

Towards answering \RQ{1}, this section examines the key determinants that define the correlation between a job's memory allocation and the resource efficiency of its execution.
Thereby, we explore the phenomenon of resource-efficient on-disk processing with in-memory data processing frameworks such as Spark.

\subsection{Workload-Dependent Memory Access}

One crucial influence on the resource usage patterns of a dataflow job is the underlying algorithm and its implementation in a distributed dataflow system.

Some algorithms iterate over a dataset only once and in a predetermined order.
Such workloads require only small memory allocations.
This class of workloads can be identified by lightweight profiling, i.e., running the job on reduced hardware and on only a sample of the data while measuring resource access patterns~\cite{will2022get}.

Another class of workloads retains access to random portions of the entire dataset throughout most of the execution.
In this case, however, the implications for resource efficiency are not straightforward.
On the one hand, allocating more memory to the cluster than what is needed to cache the entire dataset constitutes inefficient overprovisioning.
On the other hand, the impact of memory underprovisioning depends on the cost of allocating enough memory for in-memory caching relative to the resulting performance gains.
This equation is influenced by a variety of factors, the most important of which are subsequently explored.

\subsection{Job-Input-Dependent Memory Access}

Another major influence on resource usage patterns is the job input in the form of the input dataset and job parameters.
The latter can influence the behavior of the job and thus its data access patterns, with all the implications as described in the previous subsection.

\vspace{-3.0mm}
\begin{figure}[htb!]
        \includegraphics[width=.452\linewidth]{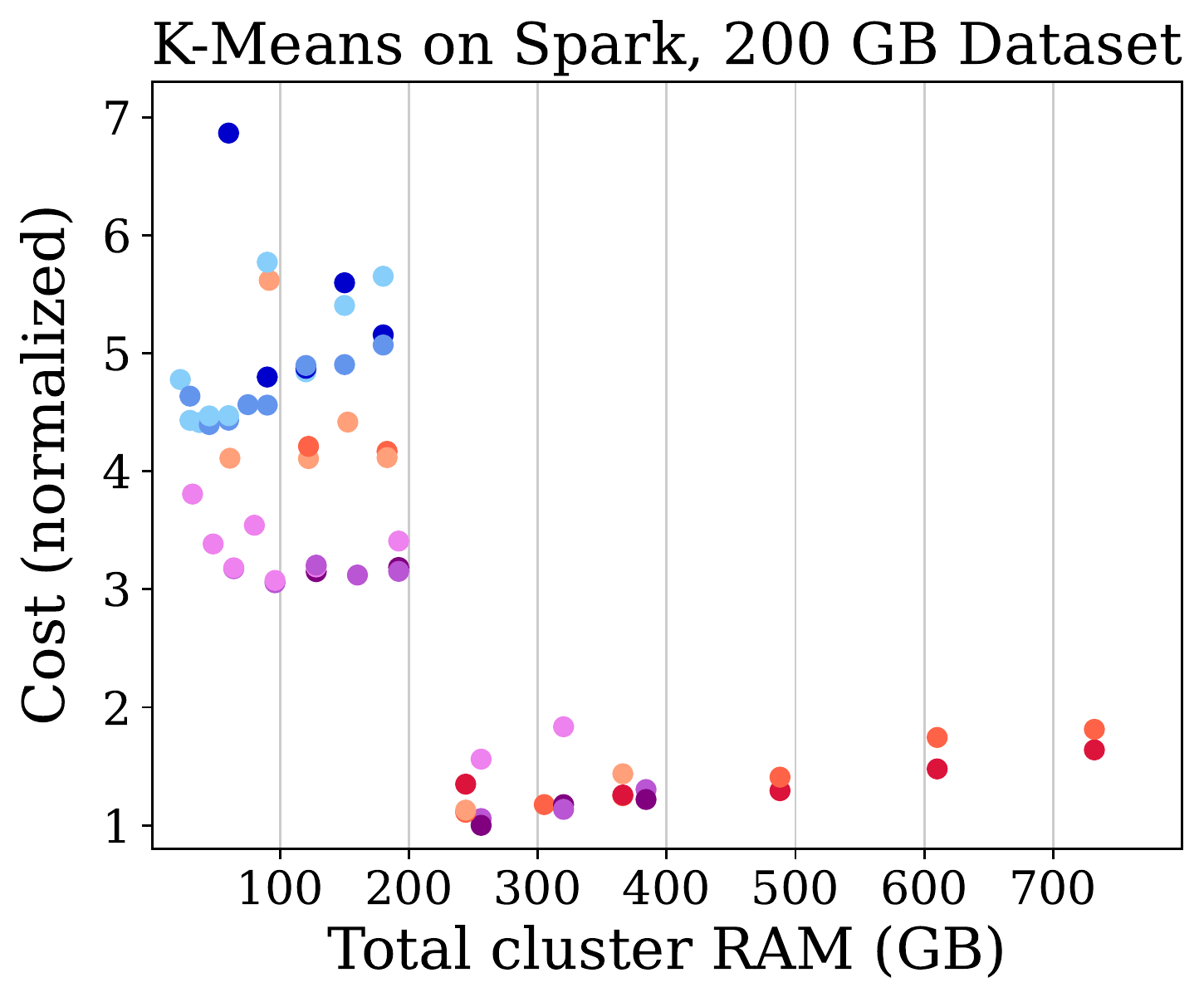}%
        \hspace{-0.655mm}%
        \includegraphics[width=.429\linewidth]{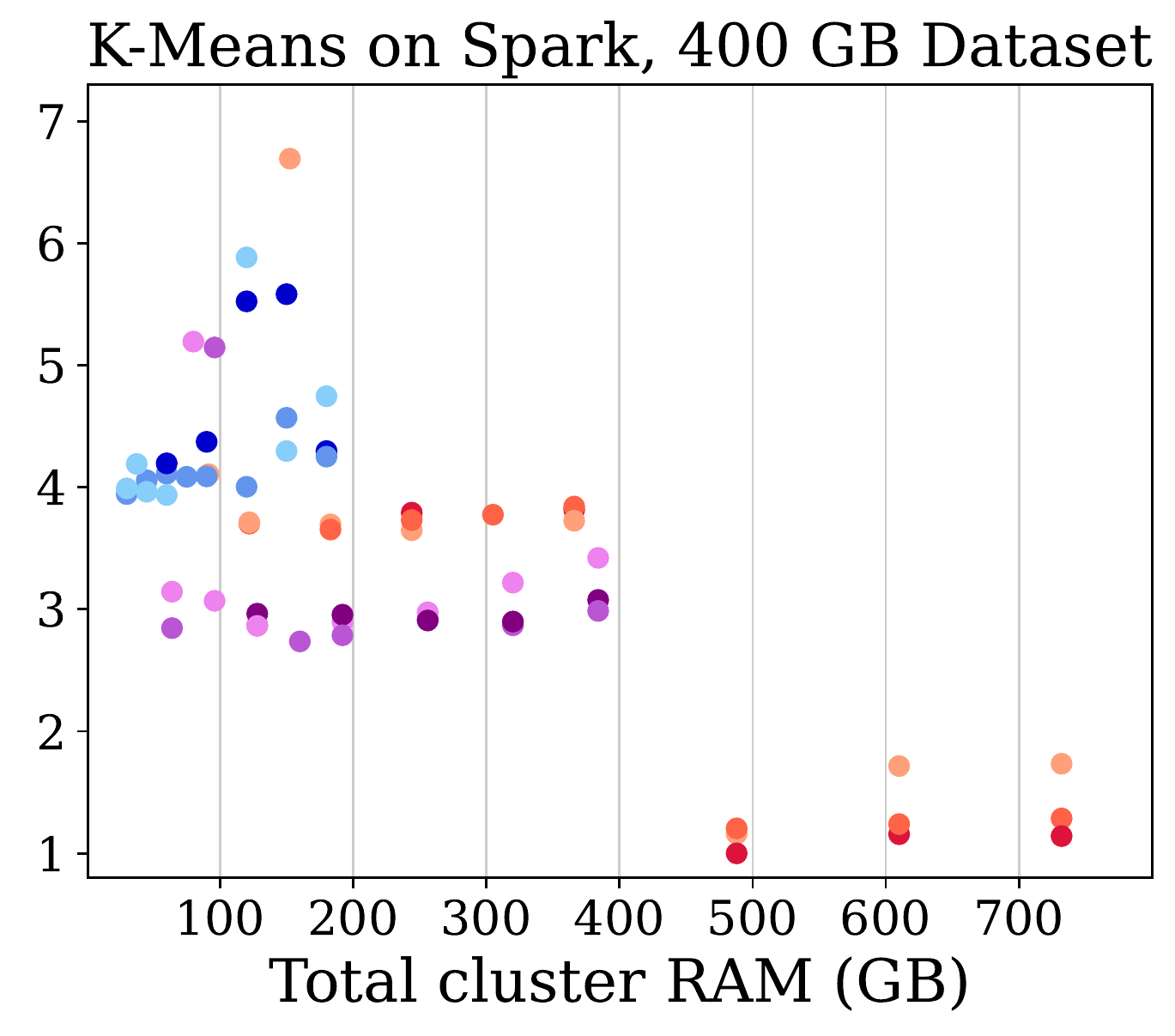}%
        \includegraphics[width=.1258\linewidth]{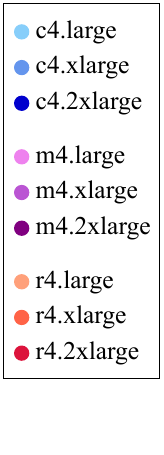}%
    \vspace{-1.5mm}
    \caption{Execution cost of K-Means on AWS in relation to allocated memory and the impact of input dataset sizes.}\label{fig:kmeans}
    \Description{Execution cost of K-Means on AWS in relation to allocated memory and the impact of input dataset sizes.}
\end{figure}

Key characteristics of the dataset, such as size, can influence how much processing is done in memory or on disk.
Figure~\ref{fig:kmeans} shows an example where the threshold for how much memory is required for cost-optimal processing shifts as the dataset size increases.
This is because K-Means iterates over a dataset multiple times and can therefore benefit greatly from fitting the entire dataset into its combined cluster memory.
At each of these iterations, caching the dataset in memory can reduce the runtime by avoiding disk I/O operations.
This reduces the amount of time that the cluster resources are reserved, which can lower the cost of the execution.

\subsection{Interdependence with Other Resource Types}

When discussing resource efficiency, the amount of memory to allocate cannot be considered in isolation.
Resource efficiency is about increasing performance by adding resources that have the lowest cost in relation to the lowered execution time.

On the one hand, this depends on the cost model, e.g., the prices of different resources in a public cloud.
On the other hand, it depends on how much the performance can be increased by these resources, e.g., how much the execution time can be reduced by adding more CPU cores to the cluster.%

\subsection{In-Memory vs. On-Disk Processing}

Figure~\ref{fig:odp} summarizes the impact of the factors described in this problem analysis by illustrating two cases: Memory bottlenecks due to underallocation versus graceful on-disk processing.

\iftrue
\vspace{-7mm}
\begin{figure}[h!]
    \centering
    \subfloat{%
      \includegraphics[width=.46\linewidth]{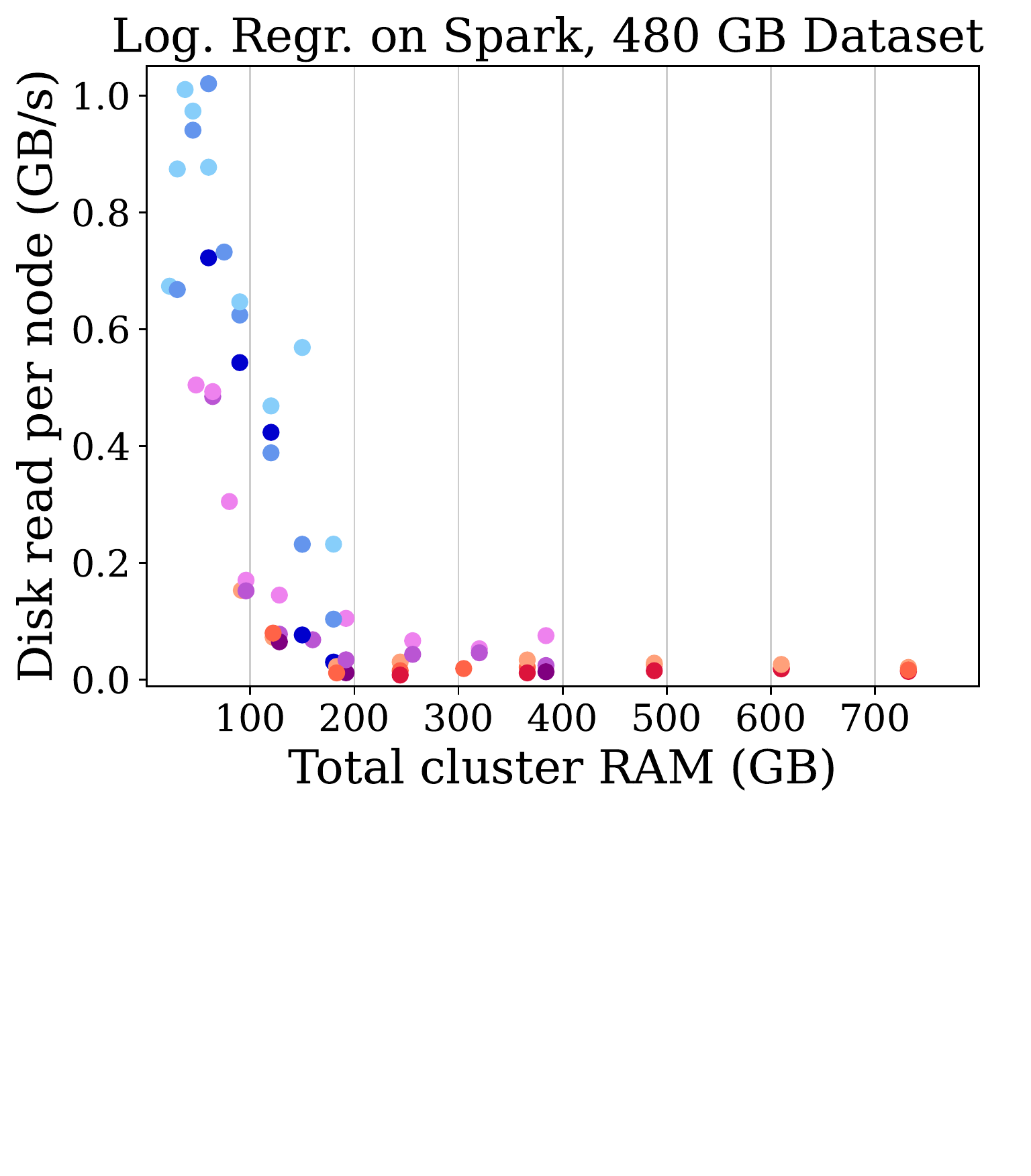}%
    }\hspace{-2.655mm}
    \subfloat{%
      \includegraphics[width=.419\linewidth]{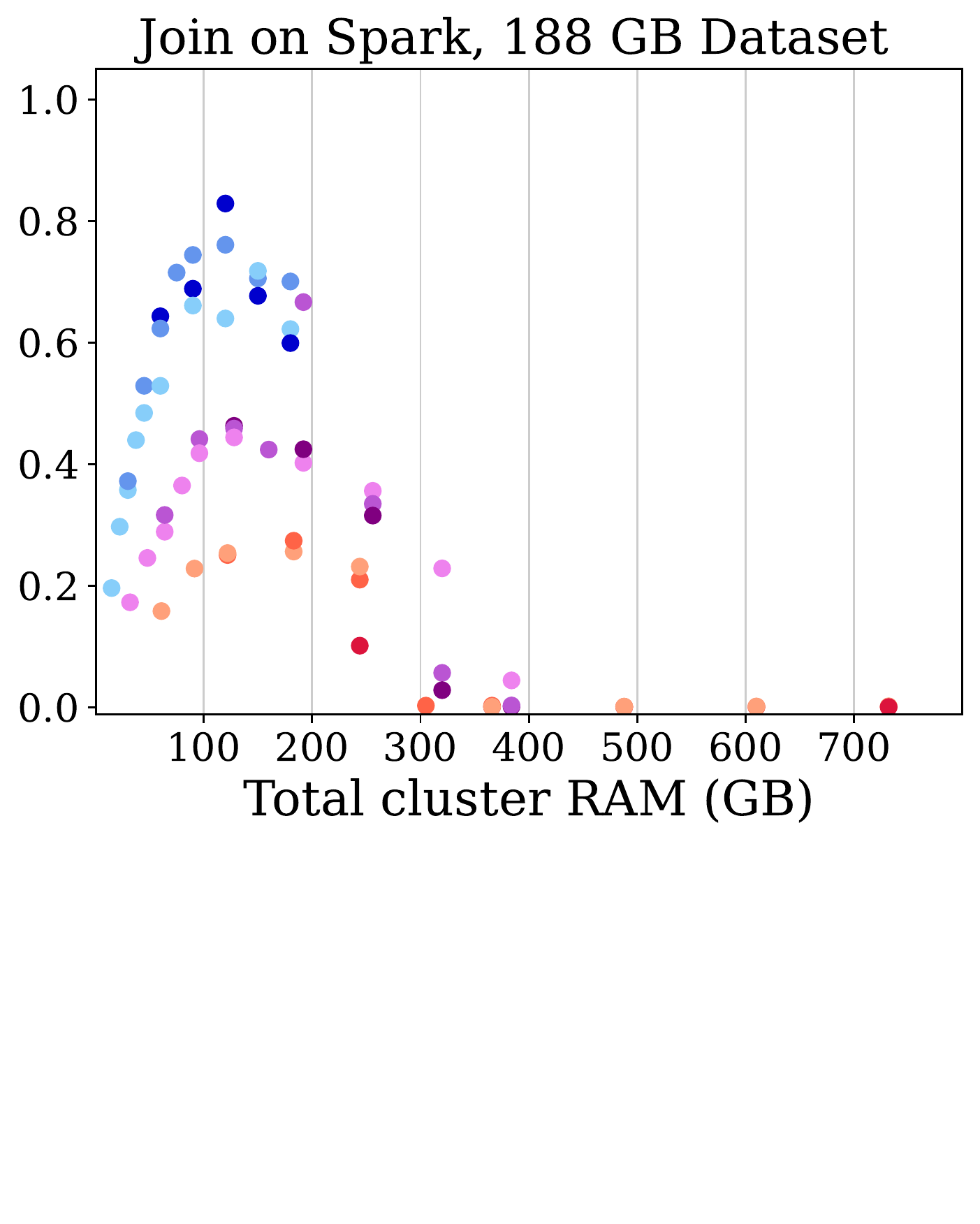}%
    }\hfill
    \subfloat{%
       \includegraphics[width=.1358\linewidth]{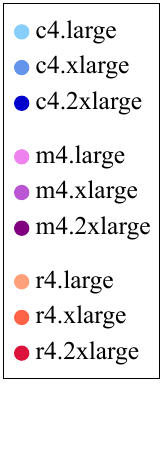}%
    }

    \vspace{-22.0mm}

    \hspace{.15mm}
    \subfloat{%
      \includegraphics[width=.433\linewidth]{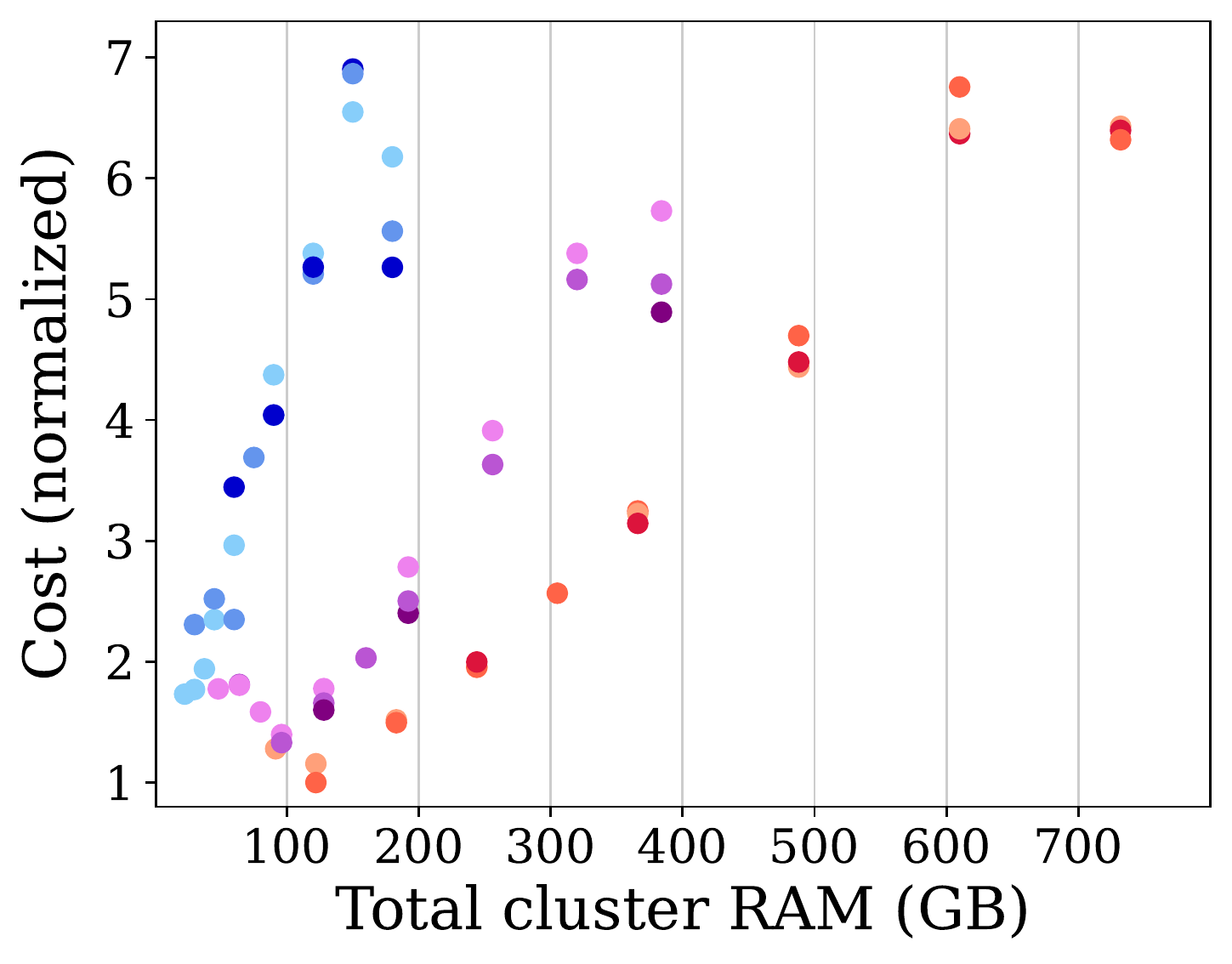}%
    }\hspace{-.45mm}
    \subfloat{%
        \includegraphics[width=.409\linewidth]{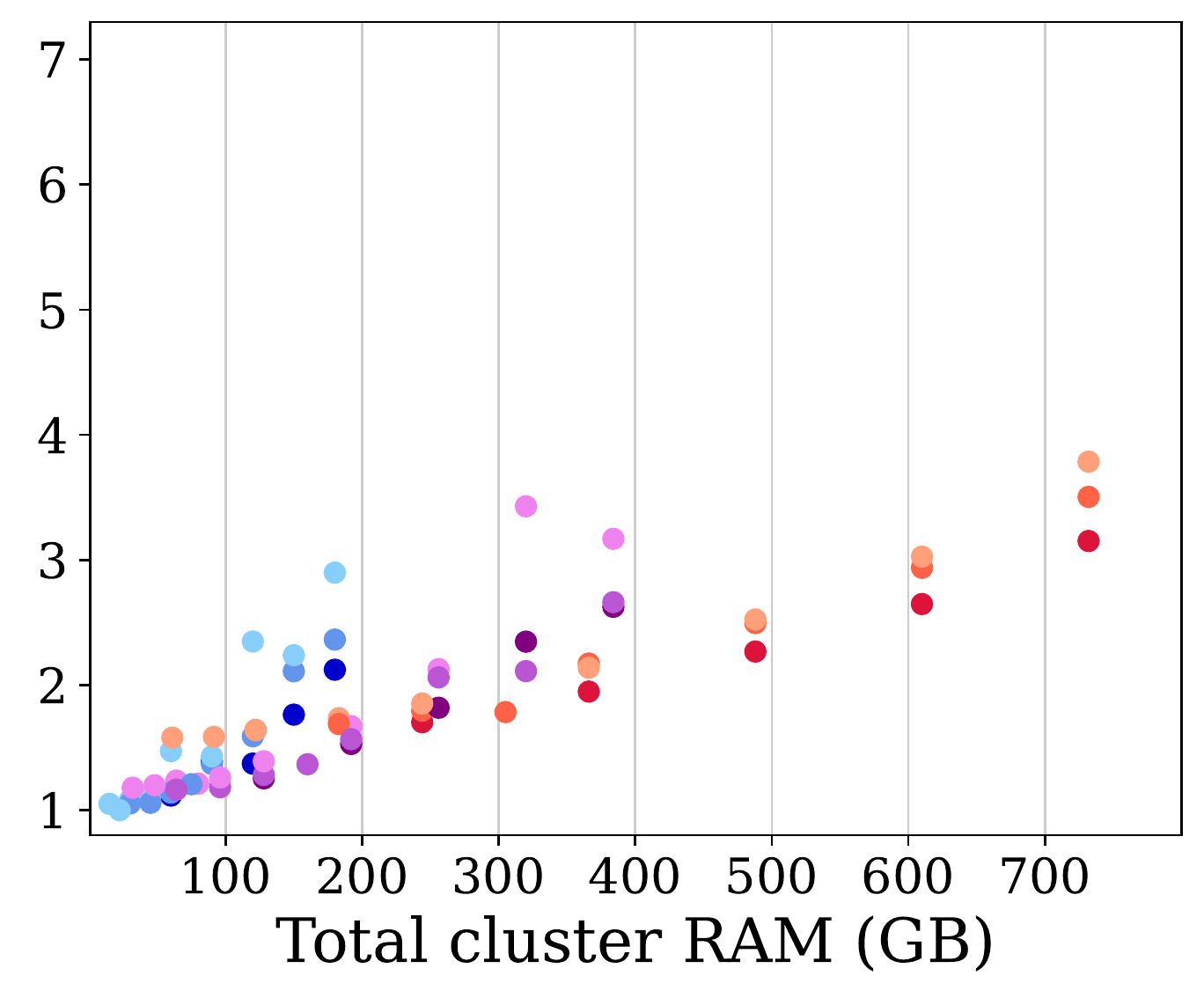}%
    }\hfill
    \vspace{-3.5mm}
    \caption{Cost-optimal in-memory processing vs.\ cost-optimal on-disk processing.}\label{fig:odp}  %
    \Description{Cost-optimal in-memory processing vs.\ cost-optimal on-disk processing.}
\end{figure}
\vspace{-5mm}
\fi

\hspace{-4.5mm}
\emph{Observation 1: From On-Disk Processing to In-Memory Processing}\\
For both jobs, we see that the disk is used when running in clusters with low total memory.
Once enough memory is available, all processing is done in memory.
For Logistic Regression with a 480~GB dataset and Join with a 188~GB dataset, this point is reached at about 170~GB and 300~GB of total cluster RAM, respectively.

While adding more memory reduces the volume of disk accesses through in-memory caching, adding more CPU cores can increase the throughput of disk accesses by speeding up the overall execution.
In the case of Join, this results in an increase in data read from the disk per second until approximately 150~GB of total cluster RAM is reached.
Thereafter, the reduced volume outweighs.

\vspace{1mm}
\hspace{-4.5mm}
\emph{Observation 2: Cost-Effectiveness of In-Memory Processing}\\
For Logistic Regression, the cost optimum is reached when about 120~GB of RAM is allocated to the cluster, at which point the job performs almost entirely in-memory processing.
For Join, the cost optimum is practically at the lowest memory and CPU allocation, despite the lack of in-memory caching in this configuration.

\vspace{1.3mm}
\hspace{-4.5mm}
For all jobs, the most cost-effective configuration is also resource-efficient.
To improve performance, this cost-effective configuration can be used as a starting point by adding the resources that accelerate the execution the most at the lowest additional cost.

\pagebreak
\section{Discussion}\label{sec:DISCUSSION}

This section discusses the results of the problem analysis in Section~\ref{sec:ANALYSIS} in terms of their implications for efficient approaches to cluster resource allocation (\RQ{2}).

\subsection{Memory Allocation}

To facilitate in-memory processing, the amount of memory that is allocated to the job depends on the size of the dataset, which can vary with each new job submission.
There are already methods that can determine the memory footprint of the dataset using lightweight profiling~\cite{al2022juggler, will2022get}.

However, whether or not in-memory processing is resource efficient is difficult to determine, since it depends on a number of interdependent factors.
While in-memory processing usually increases performance, a larger performance increase could be achieved by allocating other resources at a lower cost.
As a result, cluster resource allocation approaches require knowledge of the cost of increasing performance for all allocatable resources.

\subsection{CPU Allocation and Scaling}

\begin{sloppypar}

Allocating enough memory for in-memory processing becomes resource-efficient when the same performance gain cannot be achieved at a lower cost by adding more CPU cores instead.
Thus, eligible resource allocation approaches need to consider the scaling behavior of the job to make decisions for or against in-memory processing.
As a side benefit, this also allows users to make an informed cost-performance trade-off with their CPU allocation.
\end{sloppypar}

To complicate matters, in environments such as public clouds, the ratio of gigabytes of RAM to CPU cores is often not freely configurable, but rather resources are bundled in a VM\@.
Thus, in order to get more of one resource, the user is forced to also acquire other types of resources in conjuction, which may be underutilized, affecting overall resource efficiency.

Finally, there is the decision to scale a job horizontally or vertically, i.e., to scale by adding nodes or increasing node sizes.
In our analyzed dataset, this influence is less significant than the total amount and ratio of memory and CPU cores, as can be seen in Figure~\ref{fig:one}, Figure~\ref{fig:kmeans}, and in the lower half of Figure~\ref{fig:odp}.
If this influence does not depend on the memory-CPU balance, it may be possible to address that decision last in a resource allocation approach.

\subsection{Efficient Strategies for Resource Allocation}

Making an informed decision about resource allocation requires knowledge of the many factors that influence the cost and performance of a distributed dataflow job.
However, the cost of gathering information must not offset the efficiency gains achieved during execution.

Therefore, all relevant information must be gathered in an efficient manner, e.g., by learning certain job behaviors through lightweight profiling on reduced hardware and a representative sample of the dataset.
Furthermore, the information gathered from profiling and full executions should be reused in subsequent executions of similar jobs.
This requires an awareness of influences that are specific to the job itself, as opposed to those that must be adaptively reconsidered when the execution context changes, such as memory allocation in relation to the dataset size. %

\pagebreak
\section{Related Work}

This section discusses existing approaches to cluster resource allocation for efficient distributed data processing and highlights their main limitations.

\subsection{Black-Box Performance Models}

Several related works on cluster resource allocation build black-box performance models to learn the behavior of jobs on different cluster configurations~\cite{venkataraman2016ernest, thamsen2016bell, alipourfard2017cherrypick, hsu2018arrow, chen2021silhouette, scheinert2021potential, will2022ruya}.

Some approaches then use these models to predict the runtime and thereby also the cost of execution on the given cluster, allowing users to make trade-offs according to their preferences~\cite{venkataraman2016ernest, thamsen2016bell, chen2021silhouette, scheinert2021potential}.\\
\emph{Ernest}~\cite{venkataraman2016ernest}, a prominent example of such approaches, trains a parametric model for the scale-out behavior of jobs on the results of sample runs on reduced input data.
This strategy is viable for recurring programs that exhibit a rather intuitive scale-out behavior.
The critical drawback of an approach like Ernest is that the used models require a significant amount of training data to become viable, the generation of which causes considerable overhead.

Other related work iteratively searches for a configuration for just the given job, focusing on optimizing for particular metrics such as low cost or high execution speed~\cite{alipourfard2017cherrypick, hsu2018arrow, will2022ruya}.\\
\emph{CherryPick}~\cite{alipourfard2017cherrypick} attempts to directly predict the cost-optimal cluster resource configuration that satisfies a given runtime target by utilizing Bayesian optimization.
The search process stops when it has found the optimal configuration with reasonable confidence.
Although this process is faster than building a complete performance model, it still imposes overhead through many executions until it converges on an optimal solution.

A significant limitation common to both of these general approaches is that these models must be retrained, at least in part, whenever the execution context changes.
This could be when new hardware options become available, or when the job's inputs change, either in the form of parameters or the dataset.

\subsection{Memory-usage-aware approaches}

Recent approaches that address the limitations of costly trained performance models treat memory allocation as the key factor in resource efficiency~\cite{al2022blink, al2022juggler, will2022get, will2022ruya}.
These approaches attempt to determine the memory needs of a job through lightweight profiling of the job, focusing on facilitating in-memory processing.\\
\emph{Juggler}~\cite{al2022juggler} is an example of an approach that selects datasets to be cached for a job and estimates the cache's memory requirement in relation to the dataset size.
However, Juggler is only designed specifically for iterative machine learning jobs in Spark.

The main limitation of all these approaches at the moment is that they do not attempt to distinguish situations where in-memory caching is efficient from ones where the cost of allocating enough memory is not justified by the performance gains realized.
This remains an important problem because the approach of allocating enough memory for in-memory caching can be inefficient, even for iterative machine learning jobs.
Such situations arise, for example, when the cost of adding memory for caching is comparatively high and the cost of adding CPU cores or other resources is comparatively~low.

\section{Conclusion}\label{sec:CONCLUSION}

In summary, this paper has explored the key considerations for allocating efficient cluster resources to a distributed dataflow job, with a focus on memory.
The specific challenge we identified is the question of when in-memory processing versus on-disk processing is resource efficient.
Finally, in accordance with these preliminary results, we explored requirements for building an improved resource allocation solution that does not rely on costly trained models.

In the future, we will work on efficient methods to automatically evaluate the memory requirements \emph{and} scaling behavior of jobs.
Moreover, we hope our short paper also inspires further research by others in the same direction.

\begin{acks}
This work has been supported through a grant by the German Research Foundation (DFG) as ``C5'' (grant 506529034).
\end{acks}

\bibliographystyle{ACM-Reference-Format}
\bibliography{./references}

\end{document}